\documentclass[aps,pra,preprint,endfloats]{revtex4-1}

\usepackage{graphicx}
\usepackage{inputenc}
\usepackage{xcolor}

\begin{document}

\title{Luminescent defects in a few-layer h-BN film grown by molecular
beam epitaxy}
\author{A. Hern\'andez-M\'inguez}
\email{alberto.h.minguez@pdi-berlin.de}
\author{J. L\"ahnemann}
\author{S. Nakhaie}
\author{J. M. J. Lopes}
\author{P. V. Santos}
\affiliation{Paul-Drude-Institut f\"ur Festk\"orperelektronik, Leibniz-Institut im Forschungsverbund Berlin e.V., Hausvogteiplatz 5-7, 10117 Berlin, Germany}

\date{\today}

\begin{abstract}
We report on luminescent centers contained in a few-layer-thick hexagonal boron nitride (h-BN) film grown on Ni by molecular beam epitaxy. After transfer to a SiO$_2$/Si substrate, sharp lines are observed in photo- and cathodoluminescence spectra in both the ultraviolet and the visible range. Spatially resolved measurements reveal that the luminescent centers responsible for these lines are localized within microscopic multi-layer islands that form at the nucleation centers of the h-BN film. The comparison of their energy, polarization and phonon replica emission with previous theoretical predictions suggest that the N$_\mathrm{B}$V$_\mathrm{N}$ anti-site could be one of the light emitters present in our sample. Moreover, we have also observed evidences of other kinds of centers that could be associated to defects containing carbon or oxygen. The characterized luminescent defects could have potential applications as quantum light sources.
\end{abstract}

\maketitle

\section{Introduction}\label{sec_Intro}

The controlled generation and manipulation of single photons is a key factor for the development of quantum information technologies~\cite{OBrien_NatPhot3_687_2009}. One of the most promising platforms for the realization of single photon sources (SPSs) are atom-like emitters based on solid-state systems, since they combine the optical properties of atoms with the scalability of solid-state nanostructures~\cite{Aharonovich_NP10_631_16}. In addition to the well-known quantum dots~\cite{Michler_Nature406_968_2000} and color centers in solids~\cite{Kurtsiefer_PRL85_290_2000}, the family of solid-state SPSs includes new low-dimensional systems like carbon nanotubes~\cite{Hoegele_PRL100_217401_2008} and two-dimensional materials like WSe$_2$~\cite{Srivastava_NatNanotech10_491_2015, Koperski_NatNanotech10_503_2015}. One of the most recent additions to the last group has been hexagonal boron nitride (h-BN), where luminescent defects acting as SPSs have been reported both in the visible~\cite{Tran_NatNano11_37_2016} and in the ultraviolet (UV) spectral ranges~\cite{Bourrellier_NanoLett16_4317_2016}. 

In the last years, several groups have intensively studied the optical properties of the SPSs  emitting in the visible spectrum both in bulk h-BN~\cite{PhysRevApplied.5.034005, Martinez_PRB94_121405_16, Vuong_PRL117_097402_2016}, as well as in flakes \cite{Jungwirth_NL16_6052_16, Tran_ACSNano10_7331_2016, Chejanovsky_NanoLett16_7037_2016, Shotan_ACSPhotonics3_2490_2016, Sontheimer_PRB96_121202_17, Exarhos_ACSNano11_3328_2017, Kianinia_ACSPhot4_768_2017, Grosso_NC8_705_17, Kianinia_NatComm9_874_2018}. Although the structural nature of such centers is still under discussion~\cite{Tran_NatNano11_37_2016, Tawfik_Nanoscale9_13575_2017, Wu_PRMat1_071001_2017, Sajid_PRB97_064101_2018, Weston_PRB97_214104_2018}, their good polarization properties~\cite{Tran_NatNano11_37_2016, Jungwirth_NL16_6052_16}, and the fact that they maintain their quantum character even at elevated temperatures~\cite{Kianinia_ACSPhot4_768_2017}, makes them strong candidates as optoelectronic components in integrated quantum devices. Future applications of such quantum centers, however, will depend on (\textsl{i}) the possibility of growing h-BN over large areas with high structural and morphological quality, and (\textsl{ii}) the control over the kind of luminescent defects, as well as their spatial distribution within the h-BN film.

Among the various approaches to synthesize h-BN films, molecular beam epitaxy (MBE) is very promising as it allows the formation of h-BN on different kinds of substrates with precise control over the growth conditions. Recent results have demonstrated the feasibility of MBE for the large-area growth of h-BN films on Ni~\cite{Nakhaie_APL106_213108_2015}, as well as on pyrolytic graphite~\cite{Cheng_JVSTB36_02D103_2018} and on graphene on SiC~\cite{Heilmann_2DMaterials5_025004_2018}. Moreover, the MBE-growth of vertical heterostructure films combining h-BN and graphene layers has been demonstrated~\cite{Zuo_SciRep5_14760_2015, Wofford_SciRep7_43644_2017}. 

In this contribution, we study luminescent centers in h-BN films grown using MBE. By means of spatially-resolved electron beam and laser excitation, we have investigated the spatial distribution of these centers and characterized their light emission properties. We find that the centers are mostly localized within multi-layer islands that form at the nucleation centers of the h-BN film. We clearly identify two kinds of luminescent defects that have been recently classified as SPSs: a defect emitting in the UV range that could be associated to carbon impurities, and a defect emitting in the visible spectral range that we tentatively attribute to the N$_\mathrm{B}$V$_\mathrm{N}$ anti-site. Moreover, we also observe another source of anti-bunched photons emitting in the visible spectral range that could be related to electronic transitions within a third kind of luminescent defect.

\section{Methods}\label{sec_Methods}

The few-layer-thick h-BN used in our experiment was grown by MBE on a 300 nm-thick Ni film with an area of $10\times10$~mm$^2$ that was previously evaporated on a MgO(111) substrate at room temperature. After transferring the Ni-coated substrate through air to the MBE system, it was treated by cyclic thermal annealing and Ar sputtering in order to prepare and improve the Ni surface for the MBE growth of the boron nitride~\cite{Wofford_SciRep7_43644_2017}. During the synthesis of the h-BN film, a high temperature effusion source provided the beam of B, while active N species were produced by an RF plasma source. The average thickness of the h-BN film is one nanometer, i.e.\ about three monolayers. More details about the growth procedure are described by Nakhaie \textit{et al.}~\cite{Nakhaie_APL106_213108_2015}.

After formation of the h-BN film, we transferred an area of $4\times4$~mm$^2$ from the growth template onto a $8\times8$~mm$^2$ SiO$_2$/Si wafer using a wet chemical transfer technique~\cite{Suk_ACSNano5_6916_2011}. To this end, the h-BN was first spin-coated with PMMA, and then the sample was immersed in a shallow beaker containing dilute nitric acid (10\%) to etch the Ni film. Next, the PMMA/h-BN stack was rinsed in water and transferred onto the SiO$_2$/Si substrate, where it was baked at 150~$^\circ$C on a hot plate to remove water molecules enclosed at the h-BN/SiO$_2$ interface. Finally, the PMMA was washed away in acetone. Figure~\ref{fig_evolution}(a) shows an optical micrograph of the h-BN film after the transfer process. 

Optical characterization at room temperature was performed using a Raman spectrometer (LabRam HR Evolution) with a continuous-wave 473~nm laser as excitation source. In the photoluminescence (PL) experiments at cryogenic temperatures (6--20 K), a continuous-wave 532~nm laser was used instead. PL spectra with polarized excitation were obtained by directing the laser through a linear polarizer and a half-wave plate to rotate the polarization angle before focusing the laser beam using an objective lens. The luminescence emitted by the sample was then collected by the same objective, directed through a 550~nm long-pass filter to block the laser beam, and sent to a single grating spectrometer containing a Si-based charge-coupled device (CCD) camera at the output slit (Horiba Scientific). A linear polarizer located in front of the input slit of the spectrometer was used to analyze the polarization of the luminescence.

Photon auto-correlation measurements were performed by coupling the PL emitted by the sample to a Hanbury Brown and Twiss (HBT) setup (see Supplementary Information for a drawing of the HBT setup). The input part consists of an optical fiber containing a 50/50 beam splitter, while the two output fibers are connected to two superconducting nanowire single photon detectors (SNSPD, Single Quantum). The output signals of the SNSPDs are connected to the start and stop inputs of a time-correlated single photon counting module (PicoHarp300). A band-pass filter placed before the input fiber was used to select the emission line to be analyzed by the HBT.

Cathodoluminescence (CL) measurements were carried out in a Zeiss Ultra55 field-emission scanning electron microscope (SEM) operated at an acceleration voltage of 5~kV and fitted with a Gatan MonoCL4 detection unit as well as a helium cold stage for cryogenic measurements. The light collected by a parabolic mirror is directed into a monochromator equipped with a 300 l/mm grating. For monochromatic CL intensity maps, the light was detected using a photomultiplier tube and a large slit width of 3~mm (corresponding to a spectral bandpass of about 36~nm) was chosen. For spectrally resolved maps, the full spectrum at each measurement point was detected using a CCD camera and a slit width of 0.2~mm (2.4~nm bandpass) was used to resolve the individual emission lines. As further discussed in the Supplementary Information, the electron beam excites also the SiO$_2$ substrate, resulting in a broad emission band between 1.7--2.0~eV. This substrate emission was removed from the CL spectra presented in the main manuscript (at the laser energy used for the PL measurements, the defect luminescence in SiO$_2$ was not excited, and thus a background subtraction was not necessary). The CL measurements were carried out at a temperature of 20~K.

\section{Results and Discussion}\label{sec_Resul}

\begin{figure}
\includegraphics[width=\linewidth]{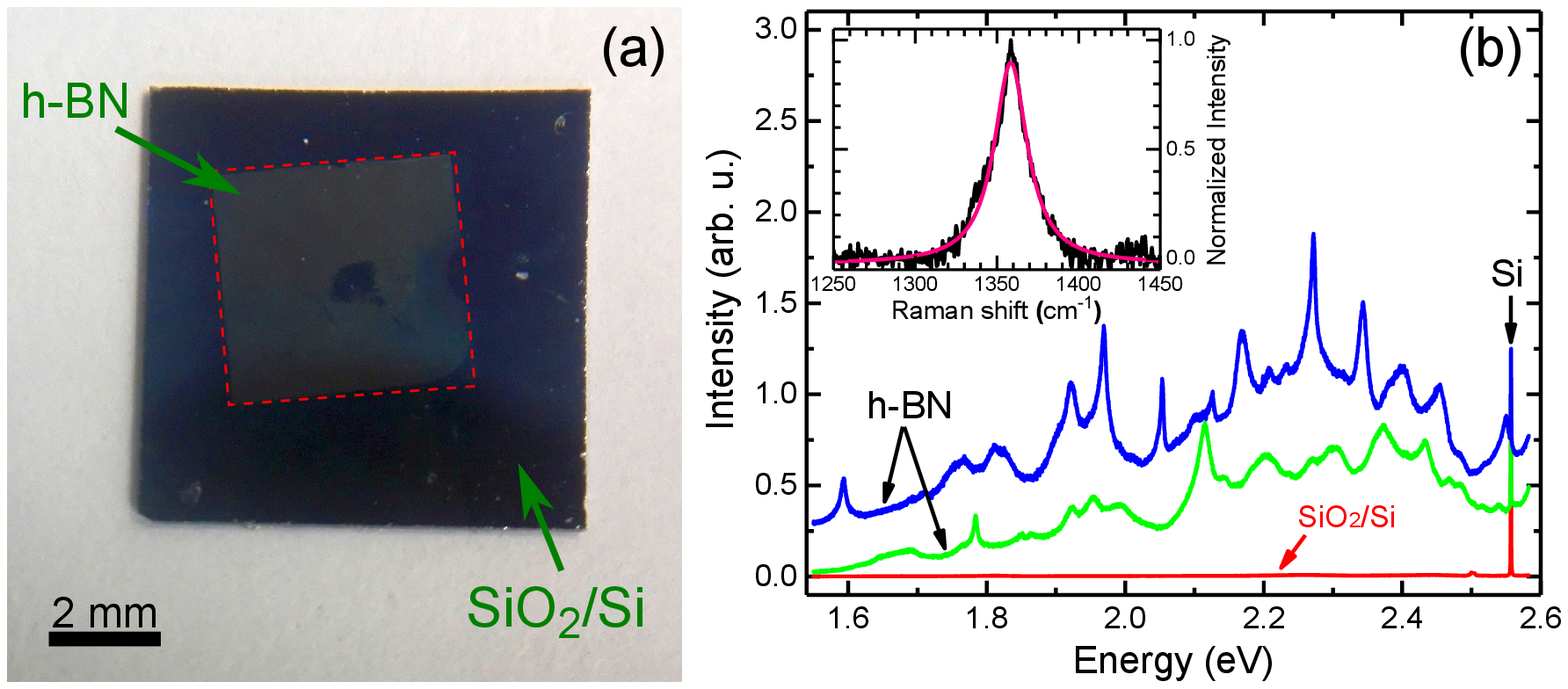}
\caption{(a) Optical micrograph of the h-BN film transferred to the SiO$_2$/Si substrate. The area covered by the h-BN film is marked by the red dashed square. (b) Room temperature photoluminescence spectra measured at two different points on the h-BN film (green and blue curves), as well as a control spectrum recorded on a region without h-BN (red curve). The curves are normalized to the amplitude of the Si Raman peak at $\approx$2.56~eV (marked in the graph) and vertically shifted for clarity. The inset displays the Raman line of the h-BN after background subtraction. The solid line is a fit to the data using a Lorentzian function.}
\label{fig_evolution}
\end{figure}

The as-grown h-BN on Ni exhibited only weak luminescence (see Supplementary Information). We attribute this to the fact that metallic surfaces may quench the luminescence of nearby photoexcited dipoles~\cite{Persson_PRB26_5409_1982, Ford_PhysRep113_195_1984}. Therefore, all experiments presented in this manuscript were performed on an h-BN film transferred to a dielectric substrate. We selected SiO$_2$/Si as substrate for the transferred film \textsl{(i)} due to the good optical contrast between h-BN and SiO$_2$, and \textsl{(ii)} to perform our luminescence measurements under equivalent experimental conditions as other authors studying these kinds of defects~\cite{Tran_ACSNano10_7331_2016, Shotan_ACSPhotonics3_2490_2016, Jungwirth_NL16_6052_16, Chejanovsky_NanoLett16_7037_2016}. Figure~\ref{fig_evolution}(b) displays room temperature PL spectra of the transferred h-BN film. The area of the sample not covered by the h-BN film (red curve) does not show any signal except for the Raman peak of the Si substrate. In contrast, the region containing h-BN exhibits a broad PL band composed of several sharp peaks. The number of peaks, their intensity and the emission energy depend on the position of the laser spot (green and blue curves). We attribute these PL spectra to the contribution of several luminescent centers contained within the area excited by the laser spot. Embedded in this PL emission, we also identified the weak Raman peak of the h-BN film at approximately 2.45~eV (not visible in the figure, see Supplementary Information for raw Raman spectra). This peak is displayed in the inset of Fig.~\ref{fig_evolution}(b) after background subtraction, and is well fitted by a Lorentzian function centered at a Raman shift of 1358.3~cm$^{-1}$.

\begin{figure}
\includegraphics[width=\linewidth]{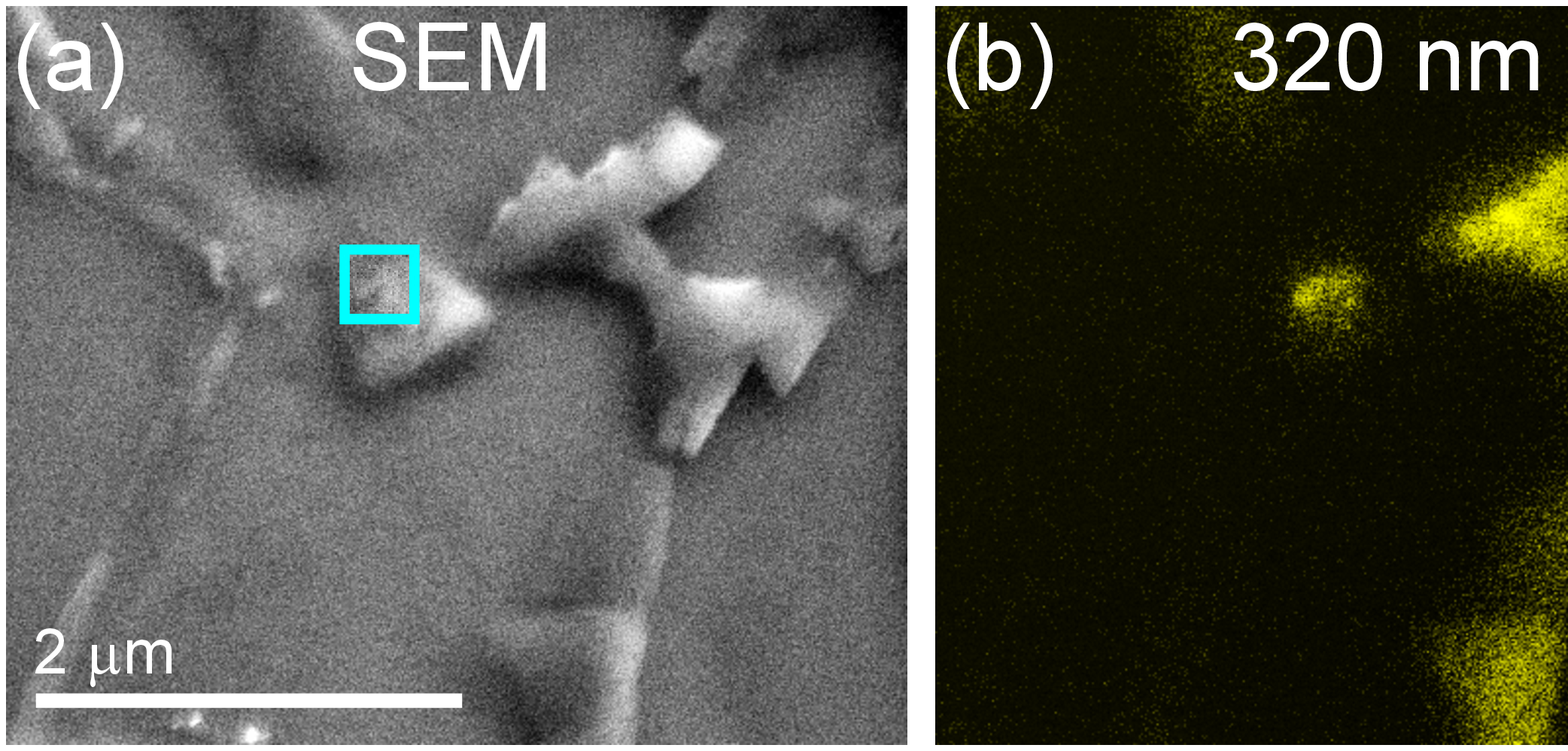}
\caption{(a) Scanning electron micrograph of a region of the h-BN film after transfer to the SiO$_2$/Si substrate. (b, c) Energy-filtered CL intensity maps centered at 320~nm (3.88~eV) and 600~nm (2.07~eV), respectively, recorded with a 36~nm spectral bandpass. The CL images are displayed in false color scale.}
\label{fig_CL2}
\end{figure}

To obtain a deeper insight into the spatial localization of the luminescent defects, we have also analyzed the h-BN film using spatially resolved CL spectroscopy. Figure~\ref{fig_CL2}(a) displays a scanning electron micrograph covering a $4.3\times3.5~\mu$m$^2$ area of the transferred h-BN film. The image shows several islands randomly distributed over the surface of the sample. They were observed across all regions of the h-BN film that we studied, thus indicating that they were homogeneously generated all over the h-BN film during the MBE growth (see Supplementary Information for a SEM image of a larger area). In fact, the islands are a few layers thicker than their surroundings and are anticipated to form around the nucleation centers~ \cite{Nakhaie_APL106_213108_2015, Heilmann_2DMaterials5_025004_2018, Xu_SciRep7_43100_2017}. In addition, atomic force microscopy confirms the presence of a significant number of ridges and wrinkles in our h-BN films~\cite{Nakhaie_APL106_213108_2015, Heilmann_2DMaterials5_025004_2018, Wofford_SciRep7_43644_2017}.

\begin{figure}
\includegraphics[width=\linewidth]{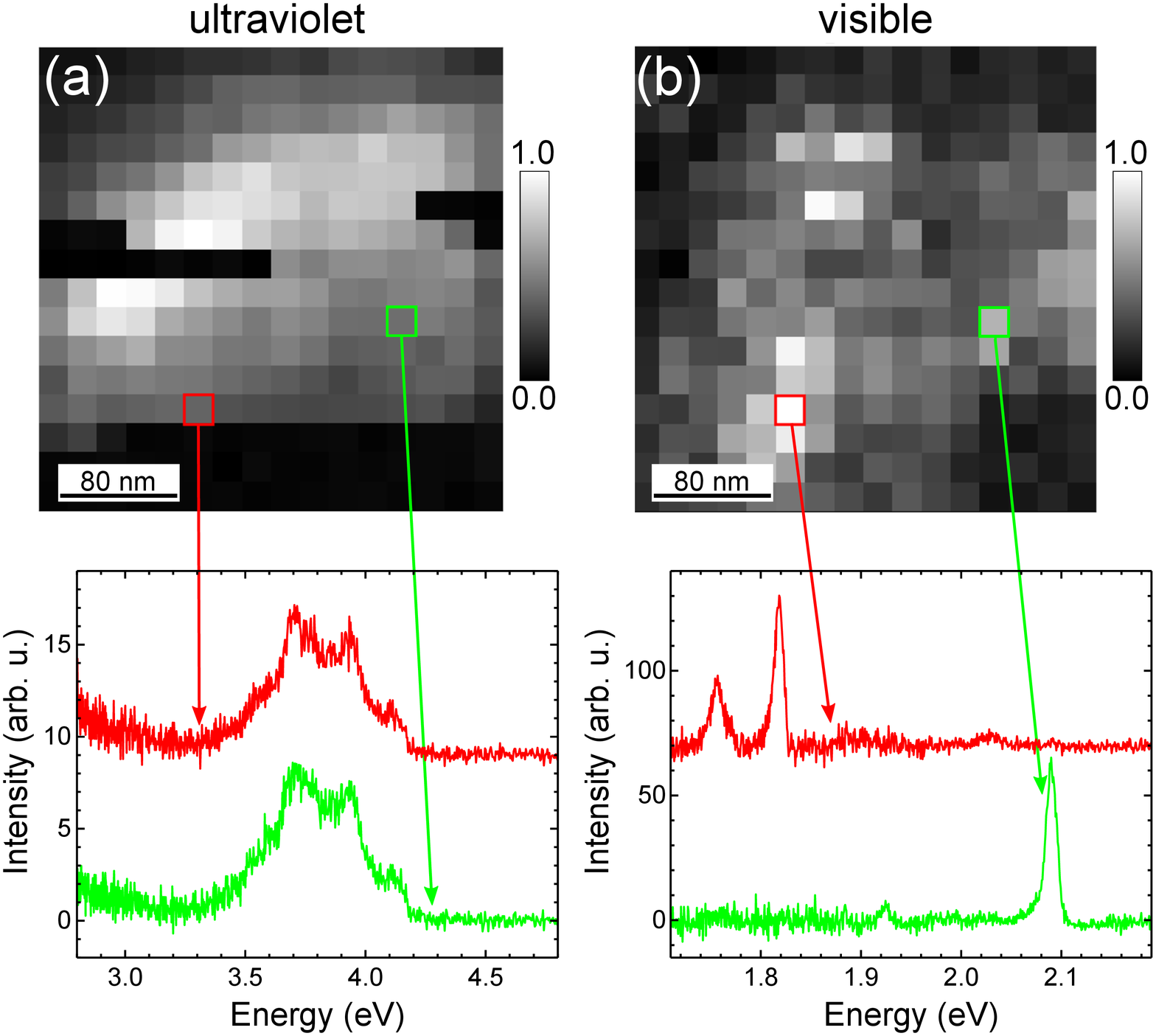}
\caption{Spatially resolved CL intensity maps of the island marked with a blue square in Fig.~\ref{fig_CL2}, for (a) the UV and (b) the visible emission range. Below the maps, exemplary spectra from two points of the map are given.}
\label{fig_CL3}
\end{figure}

While in PL the range of excited transitions is limited by the energy of the incident laser, the electron beam used for CL excites any optical transitions in the scattering volume of the beam. Figures~\ref{fig_CL2}(b) and ~\ref{fig_CL2}(c) display energy-filtered images of the low-temperature CL intensity of the same area shown in Fig.~\ref{fig_CL2}(a), collected for detection wavelengths centered at 320~nm (3.88~eV) and 600~nm (2.07~eV), respectively. These two wavelengths represent the main emission bands of the sample (see Supplementary Information for full CL spectrum). Remarkably, the images indicate that the luminescence originates from the islands rather than from the flat areas of the h-BN film. Moreover, the dominant contribution to the CL intensity comes from the centers emitting in the UV and not from those emitting in the visible range. For both wavelength regions, we have collected two-dimensional, spectrally-resolved CL maps of single islands at step sizes of 20~nm. Figure~\ref{fig_CL3} shows the maps obtained for the area marked with a blue square in Fig.~\ref{fig_CL2}(a). In the UV range (around 3.9~eV), Fig.~\ref{fig_CL3}(a) demonstrates that the CL spectrum is fairly homogeneous within the island, although with significant variations in intensity. The spectrum consists of 3 main peaks at fixed energies, separated by $\approx$180~meV from each other. A similar UV spectrum has been reported for single photon emitters contained in h-BN flakes obtained by chemical exfoliation~\cite{Bourrellier_NanoLett16_4317_2016}. These centers have been tentatively attributed to lattice defects consisting of substitutional C atoms on N sites, C$_\mathrm{N}$~\cite{Katzir_PRB11_2370_1975, Silly_PRB75_085205_2007}. However, recent theoretical calculations suggest C$_\mathrm{B}$ as a better candidate~\cite{Weston_PRB97_214104_2018}. The observation of these emission lines suggests that C impurities are present in our sample.

The spectrally-resolved map of the centers emitting in the visible range (around 2.1 eV) is displayed in Fig.~\ref{fig_CL3}(b). Contrary to the UV map, the spectrum of the centers emitting in the visible depends on the position of the electron beam within the island, thus suggesting the contribution of different kinds of localized centers. The emission of each of these centers can be seen over an area of around $100\times100$~nm$^2$, which is a result of the spatial resolution of the CL measurements being limited by the scattering volume of the electron beam in the sample and the diffusion of excited carriers within the layer.

After confirming the presence of luminescent centers in our h-BN film, we have characterized the optical properties of several centers emitting in the visible range with respect to excitation laser power and polarization. This analysis was performed by PL spectroscopy at liquid He temperatures in order to enhance their PL intensity and to reduce the linewidth of their zero phonon line (ZPL)~\cite{Jungwirth_NL16_6052_16, Sontheimer_PRB96_121202_17}. In the following, we show results for two of these centers, which we have labeled as D1 and D2.

\begin{figure}
\includegraphics[width=\linewidth]{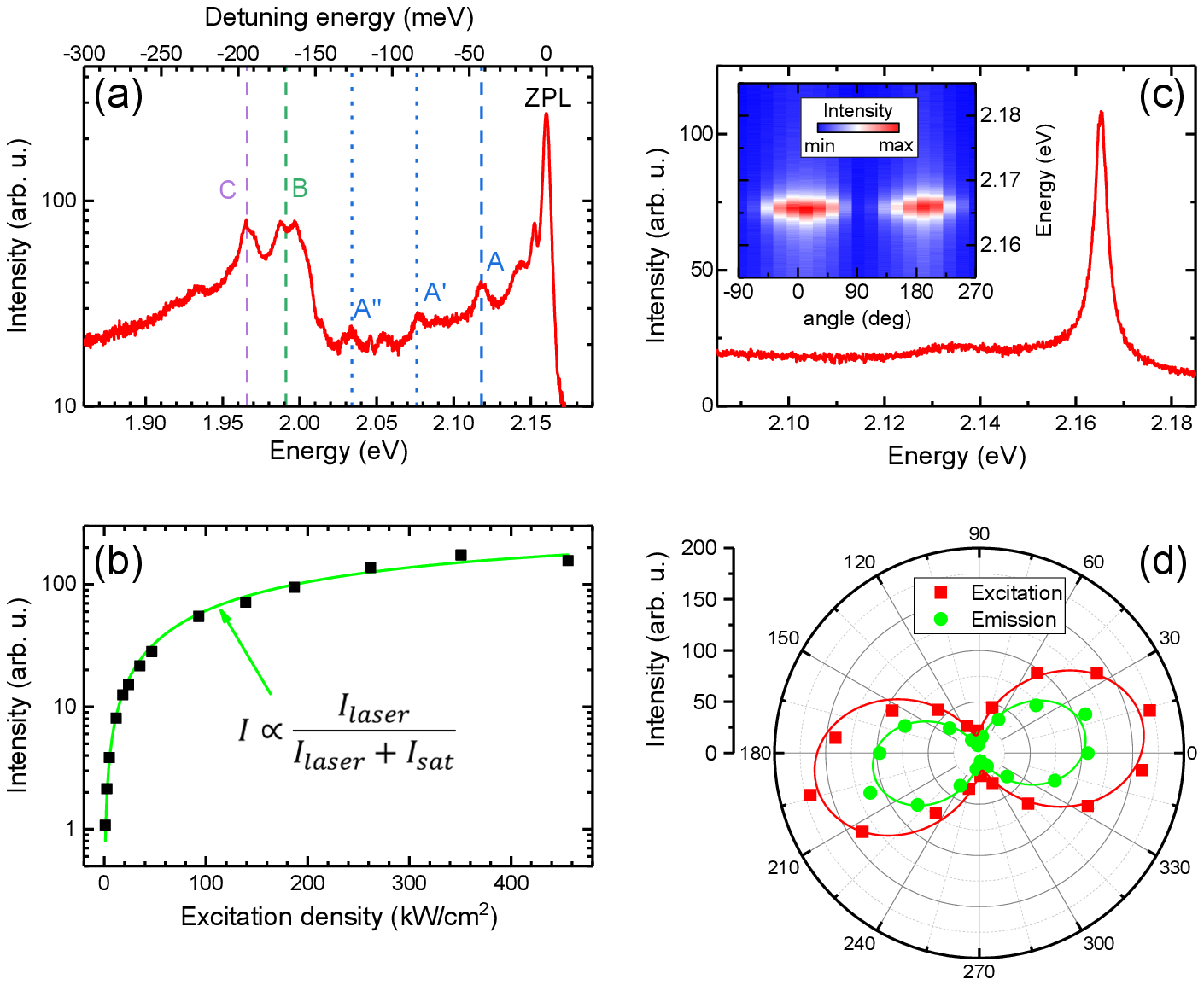}
\caption{(a) Characteristic PL spectrum of the luminescent center denoted as D1. It shows the ZPL at 2.16~eV (measured at 6~K), together with phonon side bands at lower energies associated to the vibration modes A, B and C described in the main text. The peaks marked as A' and A'' correspond to phonon replicas involving the emission of two and three phonons, respectively. (b) Intensity of the ZPL as a function of the excitation power density,  $I_{laser}$. The solid curve is a fit to the data according to Eq.~\ref{eq_IvsPlaser}. (c) Spectrally resolved PL (measured at 20~K) of a center with ZPL at 2.166~eV. The color plot of the inset displays the intensity of the ZPL as a function of the polarization angle of the excitation laser. (d) Excitation (red squares) and emission (green circles) polarization curves for the ZPL of panel (c). The red and green curves are fits using a $\sin^2\theta$ function.}
\label{fig_peakA&B}
\end{figure}

Figure~\ref{fig_peakA&B}(a) displays the PL spectrum of D1, measured at 6~K. It shows a sharp ZPL at 2.16~eV, together with a broad secondary emission with well-resolved peaks at 1.997~eV, 1.988~eV and 1.965~eV. The detuning energies of these peaks with respect to the ZPL are -163~meV, -172~meV and -195~meV, respectively. As the phonon density of states in bulk h-BN shows maxima for transverse and longitudinal optical phonons at this range of energies~\cite{Kern_PRB59_8551_99}, we attribute this secondary emission to a phonon side band (PSB) associated to electronic transitions involving the emission of one photon and one phonon. Moreover, the vibronic spectrum around the ZPL is asymmetric and consists of a low-energy side band with maximum intensities at detuning energies of about -8~meV and -16~meV. This emission resembles the one reported for the UV defect in bulk h-BN~\cite{Vuong_PRL117_097402_2016}, and therefore we attribute it to the coupling of the center with longitudinal acoustic phonons. Finally, three weak emission peaks are observed at detuning energies of about -41~meV, -82~meV and -124~meV. These peaks could be related to a vibration mode associated with the presence of the defect, as will be discussed below.

The intensity of the ZPL with respect to the excitation power density, $I_{laser}$, is displayed in Fig.~\ref{fig_peakA&B}(b). It shows the characteristic saturation behavior of this kind of systems, and is well fitted using the expression~\cite{Kurtsiefer_PRL85_290_2000}:
 
\begin{equation}\label{eq_IvsPlaser}
I\propto\frac{I_{laser}}{{I_{laser}+I_{sat}}}.
\end{equation}

\noindent Here, $I_{sat}=502$~kW/cm$^2$ is the excitation density at saturation.

The optical characteristics of center D1 are very similar to those of single photon emitters that have been tentatively associated to the anti-site complex N$_\mathrm{B}$V$_\mathrm{N}$~\cite{Tran_NatNano11_37_2016, Jungwirth_NL16_6052_16, Tran_ACSNano10_7331_2016, Grosso_NC8_705_17}. A detailed analysis of the PSB in the PL spectrum of Fig.~\ref{fig_peakA&B}(a) further supports this assumption, as we are able to identify the vibration modes that, according to recent theoretical predictions, contribute mostly to the luminescence of the N$_\mathrm{B}$V$_\mathrm{N}$ defect~\cite{Tawfik_Nanoscale9_13575_2017}. In their calculations, the authors used constrained density-functional theory (DFT) to predict the atomic geometry of the excited states for a range of potential defects in h-BN, and then they determined the strength of the electron-phonon coupling for the different vibration modes of the system, including those associated with the presence of the defects. For the N$_\mathrm{B}$V$_\mathrm{N}$ anti-site, they found that the vibration modes with the highest partial Huang-Rhys factors, and therefore with the largest electron-phonon coupling, are the ones with energies $E_\mathrm{A}=42$~meV, $E_\mathrm{B}=169$~meV and $E_\mathrm{C}=194$~meV. Mode A consists of a delocalized phonon mode in which the defect site moves along the dipole direction. Modes B and C, in contrast, are localized modes where the largest displacements originate from the atoms surrounding the defect site. In the B mode, the two boron atoms in the defect site stretch perpendicular to the dipole direction, while in the C mode the nitrogen atom in the defect stretches further into the defect site. We have marked the photon emission energies associated to the phonon modes A, B and C with vertical dashed lines in the spectrum of Fig.~\ref{fig_peakA&B}(a), obtaining an almost perfect agreement between theoretical prediction and experimental data. Moreover, in the case of mode A, we can even resolve transitions involving the emission of two and three phonons -- labeled as A$'$ and A$''$, respectively.

An additional property of the N$_\mathrm{B}$V$_\mathrm{N}$ center is the fact that its PL is polarized in the plane and along the $C_{2v}$ symmetry axis of the defect~\cite{Tran_NatNano11_37_2016, Sajid_PRB97_064101_2018, Jungwirth_NL16_6052_16}. Figures~\ref{fig_peakA&B}(c) and (d) show the PL polarization dependence of a luminescent center emitting at a similar energy as the one in Fig.~\ref{fig_peakA&B}(a), but with higher intensity. Figure~\ref{fig_peakA&B}(c) displays the ZPL (measured at 20~K), which is well fitted by a single Lorentzian function centered at 2.166~eV. The color plot of the inset shows the dependence of the PL spectrum (vertical scale) on the polarization angle of the exciting laser (horizontal scale). It shows two clear maxima and minima within one full rotation of the polarization axis. Figure~\ref{fig_peakA&B}(d) displays the intensity of the ZPL with respect to the polarization angle of the laser excitation (red squares), together with the polarization analysis of the emitted PL (green circles). Both sets of data are well fitted by a $\sin^2\theta$ function with maxima aligned along the same direction, thus indicating that the optical transition is governed by two electronic levels with a single absorption and emission dipole, in agreement with theoretical predictions~\cite{Jungwirth_PRL119_057401_2017, Sajid_PRB97_064101_2018}. 

\begin{figure}
\includegraphics[width=\linewidth]{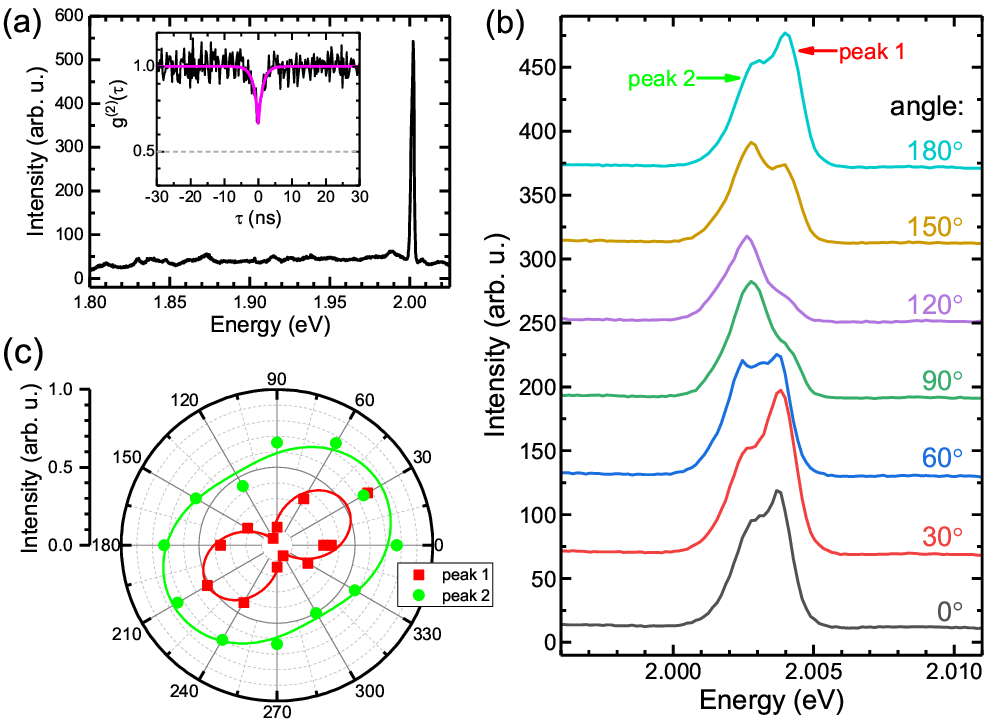}
\caption{(a) Spectrally resolved PL of the center denoted as D2. It shows the ZPL at 2.003~eV (measured at 6~K) and no traces of a phonon side band. Its second-order autocorrelation function, $g^{(2)}(\tau)$, is displayed as inset. The solid curve is a fit to the data according to Eq.~\ref{eq_g2}. (b) Luminescence as a function of the polarization angle of the excitation laser. It consists of two emission lines, indicated as peak 1 and 2 in the figure. (c) Excitation polarization curves for peak 1 (red squares) and peak 2 (green circles). The red and green lines are fits using a $\sin^2(\theta)$ function.}
\label{fig_peakC}
\end{figure}

Not all the luminescent centers that we studied followed the behavior observed for the center D1. As an example, we display in Fig.~\ref{fig_peakC} the optical characteristics of the center labeled as D2. Its low-temperature (6~K) PL spectrum, displayed in Fig.~\ref{fig_peakC}(a), consists of a ZPL at 2.003~eV, with no clear traces of a PSB. A closer examination of the ZPL, see Fig.~\ref{fig_peakC}(b), reveals that it does not consist of a single peak, as in the case of D1, but of at least two emission peaks with an energy separation of about 1.55~meV. Interestingly, each peak exhibits a different dependence on the polarization angle of the laser excitation, cf. Fig.~\ref{fig_peakC}(b) and (c). While the high energy peak (denoted as peak 1) shows a clear dipolar behavior similar to the one observed for defect D1, cf. Fig.~\ref{fig_peakA&B}(d), the low energy peak (denoted as peak 2), although aligned with the polarization axis of peak 1, is only weakly polarized.

The presence of several peaks in the ZPL is also reflected in its photon auto-correlation measurement, shown in the inset of Fig.~\ref{fig_peakC}(a). The photon correlation histogram displays a clear anti-bunching feature at zero time delay, which is well fitted by a two-level model:

\begin{equation}\label{eq_g2}
g^{(2)}(\tau)=1-a\exp(-|\tau|/\tau_R),
\end{equation}

\noindent where $\tau_R\approx1.3\pm0.2$~ns is the recombination lifetime and $g^{(2)}(0)=0.66\pm0.04$ ($0.48\pm0.05$ after background correction) is the auto-correlation value at zero time delay. According to the quantum theory of light, $g^{(2)}(0)$ is related to the number $N$ of photons simultaneously emitted by the expression $g^{(2)}(0)=1-1/N$. The experimental value of $g^{(2)}(0)$ is above the upper limit $g^{(2)}(0)<0.5$ that characterizes single photon sources, thus indicating the simultaneous emission of $2<N<3$ photons. This result suggests that the observed spectrum originates from two electronic transitions emitting single photons at slightly different energies.

In contrast to D1, we cannot associate a clear crystal structure to defect D2. On the one hand, it could consist of just two independent N$_\mathrm{B}$V$_\mathrm{N}$ defects excited simultaneously within the area covered by the laser spot. Although the energy of the ZPL of D2 is about 160~meV lower than that of D1, it has been shown that local strain can induce such a spectral tuning in the N$_\mathrm{B}$V$_\mathrm{N}$ defect~\cite{Grosso_NC8_705_17}. On the other hand, the absence of a clear PSB, and the weaker polarization of one of the two peaks, suggests a different crystallographic nature for this emission. One possibility could be the presence of defects containing elements other than B and N, like O$_\mathrm{2B}$V$_\mathrm{N}$, C$_\mathrm{N}$Si$_\mathrm{N}$V$_\mathrm{B}$ or C$_\mathrm{B}$V$_\mathrm{N}$. Theoretical calculations using DFT have predicted for such defects ZPL emission energies of 1.90~eV, 2.03~eV and 2.08~eV, respectively~\cite{Sajid_PRB97_064101_2018}, which are close to the energies discussed in Fig.~\ref{fig_peakC}. While the incorporation of Si during the growth of the h-BN film is highly unlikely, the incorporation of O or C into the h-BN lattice cannot be ruled out. In the case of C, this assumption is further supported by the UV emission centers observed in the CL measurements, which were tentatively attributed to C impurities. Therefore, a likely candidate for the luminescence features reported in Fig.~\ref{fig_peakC} could be C$_\mathrm{B}$V$_\mathrm{N}$. This attribution is in line with the low Huang-Rhys factor predicted for such a kind of defect~\cite{Tawfik_Nanoscale9_13575_2017}, which implies a weak emission of phonon side bands, as well as with the fact that one of its possible optical transitions is predicted to be polarized perpendicular to the plane of the h-BN layer~\cite{Sajid_PRB97_064101_2018}.

\section{Conclusions}\label{sec_Concl}

To summarize, we have characterized luminescent defects contained in a few-layer-thick h-BN film grown by MBE. CL measurements reveal that the luminescent centers are concentrated within microscopic islands that form around the nucleation centers of the h-BN. In addition, they are divided in two groups, depending on whether they emit in the visible or in the UV region. The UV emitters could be related to substitutional C impurities and show a very stable spectrum distributed homogeneously within the islands. In contrast, the optical features of the defects emitting in the visible spectral range depend strongly on the position of the excitation source on the island, thus suggesting the contribution of different kinds of defects. The comparison with theoretical predictions suggests that the N$_\mathrm{B}$V$_\mathrm{N}$ anti-site could be one of the defects present in our sample, and that a second kind of center emitting anti-bunched photons could be associated to defects containing C or O.

The observation of luminescent centers with potential applications as quantum light sources in h-BN grown by MBE is a promising step towards the integration of h-BN in new quantum opto-electronic devices based on the epitaxial growth of dissimilar two-dimensional materials. As an example, the epitaxial combination of h-BN and graphene \cite{Heilmann_2DMaterials5_025004_2018, Zuo_SciRep5_14760_2015, Wofford_SciRep7_43644_2017} could lead to the scalable fabrication of quantum devices, where the access to the quantum information stored in defect centers of the h-BN film is realized not only optically, but also electronically~\cite{Brenneis_NN10_135_15}.

\section*{Acknowledgements}
The authors would like to thank Oliver Brandt for a critical reading of the manuscript, as well as H.-P. Sch\"onherr, M. H\"oricke and C. Herrmann for technical support.


%

\newpage

\end{document}